\documentclass[a4paper,fleqn,usenatbib,useAMS]{mnras}

\usepackage{hyperref}
\hypersetup{final}
\usepackage{graphicx}	
\usepackage{amsmath}	
\usepackage{amssymb}	

\usepackage[T1]{fontenc}
\usepackage{ae,aecompl}

\usepackage[varg]{newtxmath}
\usepackage{newtxtext}

\title[AGN feedback in dwarf galaxies?]{AGN feedback in dwarf galaxies?}

\author[G. Dashyan et al.]{
Gohar Dashyan$^{1}$\thanks{E-mail: dashyan@iap.fr},
Joseph Silk$^{1,2,3,4}$,
Gary A. Mamon$^{1}$,
Yohan Dubois$^{1}$,
Tilman Hartwig$^{1}$
\\
$^{1}$ Institut d'Astrophysique de Paris (UMR 7095: CNRS and UPMC - Sorbonne Universit\'es), 98 bis bd Arago, F-75014 Paris, France\\
$^{2}$Laboratoire AIM-Paris-Saclay, CEA/DSM/IRFU, CNRS, Univ. Paris VII, F-91191 Gif-sur-Yvette, France\\
$^{3}$Department of Physics and Astronomy, The Johns Hopkins University Homewood Campus, Baltimore, MD 21218, USA\\
$^{4}$ BIPAC, Department of Physics, University of Oxford, Keble Road, Oxford OX1 3RH, UK
}
\date{}

\pubyear{2017}

\begin{document}
\label{firstpage}
\pagerange{\pageref{firstpage}--\pageref{lastpage}}
\maketitle

\begin{abstract}
Dwarf galaxy anomalies, such as their abundance and cusp-core problems, remain a prime challenge in our understanding of galaxy formation. The inclusion of baryonic physics could potentially solve these issues, but the efficiency of stellar feedback is still controversial.
We analytically explore the possibility of feedback from Active Galactic Nuclei (AGN) in dwarf galaxies and compare AGN and supernova (SN) feedback. 
We assume the presence of an intermediate mass black hole within low mass galaxies and standard scaling relations between the relevant physical quantities. We model the propagation and properties of the outflow and explore the critical condition for global gas ejection. Performing the same calculation for SNe, we compare the ability of AGN and SNe to drive gas out of galaxies. We find that a critical halo mass exists below which AGN feedback can remove gas from the host halo and that the critical halo mass for AGN is greater than the equivalent for SNe in a significant part of the parameter space, suggesting that AGN could provide an alternative and more successful source of negative feedback than SNe, even in the most massive dwarf galaxies.
\end{abstract}





\section{Introduction}

In the cold dark matter (CDM) cosmological model, larger structures form through successive mergers. Therefore, dwarf galaxies ($M_{\rm vir} < 10^{11} \rm{M}_{\odot}$) are potentially left over buildings blocks of galaxies and provide a test bed for the CDM model, as the smallest probes of cosmological structure formation. The $\Lambda$CDM model has proven successful at reproducing the large scale universe, however, disparities exist between the theory and observations on small scales: the model predicts too many small galaxies (the "missing satellites" problem, \citealt{Moore99}), cuspy dark matter profiles that are not yet convincingly observed \citep{Oh11}, and the most massive dwarfs predicted by $\Lambda$CDM simulations are rarely observed (the "too big to fail" problem, \citealt{BK11}). Baryonic feedback, especially from supernovae (SNe), is a currently controversial solution to all these difficulties. Ram pressure stripping, tidal stripping, and harassment, are additional mechanisms that should occur in the group environment. It is, however, still unclear whether these mechanisms can reconcile theory and observations at the low mass end of the galaxy luminosity function. The role of SN feedback is uncertain since SNe might fail in multiphase interstellar medium (ISM) \citep{BH15}. Moreover, massive dwarf galaxies seem to require stronger feedback than SNe can provide \citep{GK13}. The role of environmental physics is also uncertain, since dwarf galaxy disagreements with the standard model seems to extend to regions where environmental effects should be small (\citealt{GK14}).

X-ray observations indicate that Active Galactic Nuclei (AGN) are present in roughly 1\% of dwarf galaxies (\citealt{Pardo2016}; \citealt{Baldassare2016}), which, combined with any plausible duty cycle, suggests a larger occupation fraction for IMBH (\citealt{Miller15}). Moreover, AGN feedback could potentially provide a unified answer to dwarf galaxy issues in the standard model \citep{Silk_feedback}.

 In this paper, we explore the possibility of AGN feedback from an intermediate mass black hole (IMBH) in dwarf galaxies. We compute the critical halo mass for gas expulsion out of the halo by the AGN outflow, following the example of \cite{DS86} for SNe. We examine how that critical mass depends on parameters, using standard scaling relations and a spherical model. This allows us to compare the roles of SNe and AGN in expelling the gas. We are considering the competition between gas retention from the gravitational force of the halo and gas expulsion by the AGN. This is a different approach from the one yielding a high mass break \citep{Bower2006,Croton2006}, which involves the role of AGN in suppressing cooling flows onto massive galaxies, thereby modifying the bright end of the galaxy luminosity function.

In Section \ref{model}, we explain the scaling relations and the physics of the outflow driven by the AGN wind. In Section \ref{implications}, we compute the critical conditions and discuss their dependence on the parameters as well as the effect of cooling. We summarize and discuss the results in Section \ref{discussion}.

\section{Model}
\label{model}
We study gas ejection from a spherical galaxy halo driven by the AGN outflow by following the propagation of the swept-up ISM (Sections \ref{structure} and \ref{propagation}), and comparing the outflow and local escape velocities. We use scaling relations to estimate the physical quantities (Section \ref{scaling}), with the aim of exploring the parameter space.

\subsection{Scaling relations}
\label{scaling}
For a given halo mass $M_{\rm{halo}}$, the parameter space to be explored consists of: the redshift, the IMBH mass, the Eddington ratio of the AGN, the fraction of mass in gas, the gas and dark matter density profiles, the velocity of the inner wind, and the lifetime of the AGN. We investigate AGN feedback in early dwarf evolution, and therefore the relevant redshift is the typical redshift of halo formation, when the typical density fluctuation corresponds to a halo mass that satisfies $\sigma(M_{\rm halo},z)=1$. We use a fitting function for $\sigma(M_{\rm halo},z)$ (app. A of \citealt{VDB02}). Therefore, the physical quantities associated to $M_{\rm halo}$ are computed at $z(M_{\rm halo})$, and varying $M_{\rm halo}$ means varying the redshift. For a given halo mass, the virial radius $R_{\rm vir}$ and velocity $V_{\rm vir}$ are uniquely defined

\begin{subequations}
\begin{align}
\label{z}
 z&=z_{\rm{ non linear}}(M_{\rm{ halo}})\ ,\\
    M_{\rm{ halo}}&= \frac{4 \pi}{3} \Delta_{\rm{ c}}(z)\rho_{\rm{ c}}E^2(z) R_{\rm{ vir}}^3\ , \\
    V_{\rm{ vir}}&= \sqrt{ \frac{GM_{\rm{ halo}}}{R_{\rm{ vir}} }}\ ,
    \end{align}
\end{subequations}

\noindent
where $G$ is the gravitational constant, $\rho_{\rm{ c}}$  is the critical density of the Universe, and  $\Delta_{\rm{ c}}(z)$ is the density contrast relative to critical, taken from \cite{BN98}, and $E^2(z)=[H(z)/H_{\rm 0}]^2$. We use a constant gas fraction

\begin{equation}
    M_{\rm{ g}} = f_{\rm{ g}} M_{\rm{ halo}}\ ,
\end{equation}

\noindent
where $f_{\rm{ g}}=\Omega_{\rm b}/\Omega_{\rm m}\simeq 0.17$. The dark matter and gas both follow the NFW profile of \cite{NFW}

\begin{subequations}
\begin{align}
    \rho_{\rm{ DM}}(r)&={\frac {\rho _{\rm{ 0}}}{{r/R_{\rm{ s}}}\left(1+{r/R_{\rm{ s}}}\right)^{2}}}\ , \\
    \rho_{\rm{ g}}& = \frac{f_{\rm{ g}}}{1-f_{\rm{ g}}} \rho_{\rm{ DM}}\ ,
    \end{align}
\end{subequations}

\noindent
where $R_{\rm{ s}}\approx R_{\rm{ vir}} / c(M_{\rm halo},z)$, and $c(M_{\rm halo},z)$ is the NFW concentration given by \cite{DM14}. The escape velocity at the virial radius is 

\begin{equation}
V_{\rm esc} =  V_{\rm{ vir}}\sqrt{2 \log{(1 + c)}/(\log{(1 + c)} - c/(1 + c))}
\end{equation}

\noindent \citep{CL96}. Using an $M_{\rm BH}-\sigma$ type relation for the black hole (BH), we estimate the BH mass

\begin{equation}
    M_{\rm{ BH}}=A \sigma^{\alpha}\ ,
\end{equation}

\noindent
where the velocity dispersion $\sigma \simeq 0.7V_{\rm vir}$ for NFW models \citep{LM01}. We vary $\alpha$ between 3.5 and 5, and for each $\alpha$ we fit the normalization $A$ to the observational data of \cite{FM2000}, \cite{Gebhardt2000}, \cite{Tremaine2000}, \cite{Gultekin2009}, and \cite{KH13} . The mechanical luminosity of the IMBH is a function of the velocity of the AGN inner wind $v_{\rm w }$, the luminosity of the AGN, given itself by the Eddington ratio $\chi$ and the Eddington luminosity,

\begin{equation}
\label{Ledd}
L_{\rm{ m}}=\frac{v_{\rm w}}{2c} L_{\rm{ AGN}} = \frac{v_{\rm w}}{2c}  \chi L_{\rm{ Eddington}} =\frac{v_{\rm w}}{2c} \chi \frac{4 \pi G c M_{\rm{  BH}} m_{\rm p}}{\sigma_{\rm T}}\ ,
\end{equation}

\noindent where $ m_{\rm p}$ is the proton mass and ${\sigma_{\rm T}}$ the Thomson cross section. Equation \eqref{Ledd} assumes that the outflow momentum flux is comparable to that
in the emitted radiation field, $L_{\rm AGN}/c$. We assume $v_{\rm w}=0.1 c$, where $c$ is the speed of light, following observations \citep{Tombesi2010, Gofford2013, KP15}. Finally, we define $t_{\rm AGN}$ as the lifetime of the AGN.

\subsection{Structure of the outflow}
\label{structure}
\begin{figure}
    \centering
    \includegraphics[width=\columnwidth]{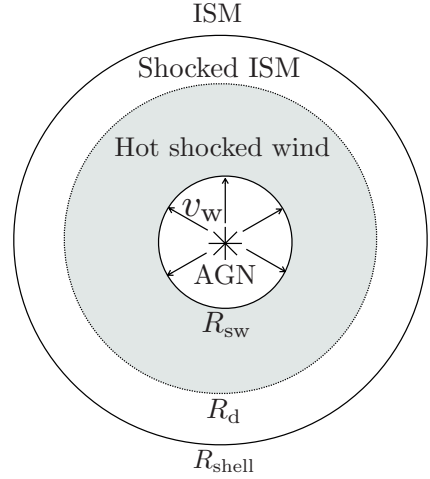}
    \caption{Schematic view of the outflow structure. The accreting IMBH drives a wind with velocity $v_{\rm w}$. It collides
with the ISM and is slowed in a strong shock at $R_{\rm sw}$. A forward shock, at $R_{\rm shell}$, is driven into
the ISM. $R_{\rm d}$ is the contact discontinuity between the shocked wind and the shocked ISM.}
    \label{cartoon}
\end{figure}

A schematic view of the outflow structure is shown in Fig.~\ref{cartoon}. The outflow from the AGN impacts the ISM of the host galaxy, producing an inner reverse shock at $R_{\rm sw}$ slowing the wind, and an outer forward shock accelerating the swept-up gas at $R_{\rm shell}$. A contact discontinuity separates the hot shocked wind and the shocked ISM at $R_{\rm d}$. The shocked wind is much hotter than the shocked ISM. Therefore, the cooling of the shocked ISM has no substantial impact on the propagation. Hence, we assume, as \cite{FG2012} that the shocked ISM is collapsed into a thin shell and define $ R \equiv R_{\rm{ d}} \approx R_{\rm{ shell}}$. Moreover, the sound crossing time in the shocked wind is smaller than the age of the outflow and the entire region is at uniform pressure \citep{weaver77}.

\subsection{Equations of the propagation}
\label{propagation}

\subsubsection{Energy-driven}

The propagation is called energy-driven when no energy is lost to radiation and the energy is thereby conserved. In that case, the motion of the shell is driven by the internal energy of the shocked wind that expands adiabatically.
Following an approach similar to \cite{weaver77} and \cite{KM92}, the set of equations giving the radius of the shell and the internal energy of the gas is

\begin{equation}
    \left \{
\begin{array}{r @{ = } l}
\label{system}
     \frac{\rm{d}}{\rm{d}t}(M_{\rm{ shell}}(R)\dot{R}R)\;\; &\;\; 2E_{\rm{ th}} + M_{\rm{ shell}} \dot{R}^2 + E_{\rm{ grav,t}}\ , \\
    E_{\rm{ in}} + E_{\rm{ grav,0}}
\;\; & \;\; E_{\rm{ th}} +  \frac{1}{2}M_{\rm{ shell}}\dot{R}^2 + E_{\rm{ grav,t}}\ , \\
\end{array}
\right.
\end{equation}

\noindent where $M_{\rm shell}$ is the mass of the gas engulfed by the blast wave at $R$, $E_{\rm{ in}}(t) =L_{\rm{ m}} \times \min(t,t_{\rm{AGN}})$ is the total injected energy; $E_{\rm{ grav,0}}$ is the gravitational energy that the swept-up gas would have had in the absence of the blast wave; $E_{\rm{ grav,t}}$ is the gravitational energy of the gas shell, under the gravity of the total mass engulfed by the blast wave. The first equation in \eqref{system} is the generalized virial theorem, applied to the gas engulfed by the blast wave, assuming that most of the mass is carried by the shell of shocked ISM. The second equation simply states the conservation of energy since the propagation is adiabatic: all the input energy ($E_{\rm{ in}}$) and the initial gravitational energy ($E_{\rm{ grav,0}}$) at a given time during propagation go to: (1) the internal energy of the shocked wind ($E_{\rm{ th}}$), (2) the kinetic energy of the shell ($\frac{1}{2}M_{\rm{ shell}}\dot{R}^2$) of the blast wave, (3) the current gravitational energy of the gas shell under the gravity of the total mass engulfed by the blast wave ($E_{\rm{ grav,t}}$).

\subsubsection{Momentum-driven}

The nature of the outflow depends on that of the reverse shock between the wind and the shocked wind, which in turn depends on the cooling: if the cooling is efficient, the shocked wind loses its thermal energy and compresses into a thin shell. The outflow is called momentum-driven because the shock is accelerated by the momentum input per unit time $L_{\rm AGN}/c$ and no longer by the adiabatic expansion of the shocked wind. The equation of propagation is

\begin{equation}
\label{momentum}
     \frac{\rm{d}}{\rm{d}t}\left[M_{\rm{ shell}}(R)\dot{R}\right] = \frac{L_{\rm AGN}}{c} -\frac{G M_{\rm shell}(R) \left[M_{\rm DM}(<R)+M_{\rm BH}\right]}{R^2}\ ,
\end{equation}

\noindent where $M_{\rm{ shell}}(R)\dot{R}$ is the momentum of the shell, and $L_{\rm AGN}/c$ is the momentum input from the photons per unit time. In the momentum-driven regime, the outflow is much less powerful than in the energy-driven regime: for an IMBH near the $M_{\rm BH}-\sigma$ relation, only a fraction $\sim \sigma/c$ of the mechanical luminosity is transferred to the ISM (\citealt{KP15}).

\subsection{Energetics}

The combined action of many SNe leads to the development of an expanding superbubble capable of sweeping-up ISM \citep{KO17}. In this Section and in Section \ref{precise} we compare the effects of AGN and SNe as a function of the halo mass, varying relevant parameters such as the Eddington ratio, the lifetime of the AGN. The star formation rate can be approximated as $M_{\rm SF} / t_{\rm SF}$  where $M_{\rm SF} \simeq M_{\rm g}(<R_{\rm vir}/10) $ is the gas mass available for star formation and $t_{\rm SF}$ is the gas depletion time in the ISM with a redshift dependance

\begin{equation}
t_{\rm SF} = 1.26\, (1+z)^{-0.34} \, \rm{Gyr} \, ,
\end{equation}

 \noindent observed by \citeauthor{Genzel2015} (\citeyear{Genzel2015}, their table 3). The total luminosity is

\begin{equation}
  L_{\rm{SN}} = E_{\rm SN} M_{\rm SF} \nu / t_{\rm SF}\ ,
\end{equation}

\noindent where $\nu$ is the number of SNe per mass of forming stars. For a Kroupa initial mass function \citep{Kroupa}, $\nu=1/150\rm{M}_{\odot}$. However, only a fraction $\epsilon_{\rm w} \sim 1-10\%$ of $L_{\rm{SN}}$  contributes to drive an outflow \citep{DT08}.

\begin{figure}
    \centering
    \includegraphics[width=\columnwidth]{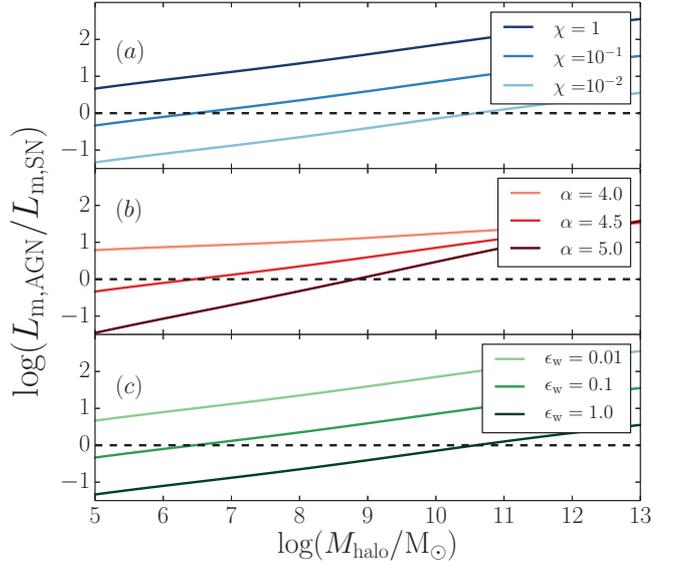}
    \caption{Ratio between the mechanical luminosity from the AGN and the mechanical luminosity from SNe, as a function of the halo mass, for different values of \textbf{(a)} the Eddington ratio $\chi$; \textbf{(b)} the exponent $\alpha$ in the $M_{\rm BH}-\sigma$ relation; \textbf{(c)} SNe wind efficiency $\epsilon_{\rm w}$. For each panel, the default values for the parameters that do not vary are: $\chi=0.1$; $\alpha=4.5$, $\epsilon_{\rm w}=0.1$.}
    \label{ratio}
\end{figure}

We compare the input wind luminosities from AGN ($L_{\rm m,AGN}$) and SNe ($L_{\rm m,SN}$) through the ratio $ \mathcal{R}=L_{\rm m,AGN}/L_{\rm m,SN}$. Fig.~\ref{ratio} shows that $\mathcal{R}$ increases with halo mass. Therefore, a halo mass exists above which the mechanical luminosity of AGN is higher than the mechanical luminosity from SNe, and this mass depends on $\chi$, $\epsilon_{\rm w}$ and $\alpha$. $\mathcal{R}$ increases with $\chi$ and decreases with $\alpha$ and $\epsilon_{\rm w}$. In Section \ref{precise}, we also take into account the momentum injection by SNe, following \cite{KO15}.

In this paper, the AGN mass is uniquely defined by the halo mass and the exponent in the $M_{\rm BH} - \sigma$ relation. By releasing that constraint, one can compute the minimum black hole mass required to energetically dominate SNe. We find

\begin{equation}
\label{mbhcrit}
\log\left( \frac{M_{\rm BH,crit}}{\rm{M}_{\odot}} \right) \simeq 2.8 + 1.1\log \left(\frac{M_{\rm halo}}{10^9\rm{M}_{\odot}} \right) + \log\left(\frac{\epsilon_{\rm w}}{\chi}\right)\, .
\end{equation}

However, this energetic approach is only first order because the energy cannot be considered as the deciding quantity a priori, since it can be lost through cooling processes. Therefore, it is important to use a self-consistent treatment of the cooling and to compute the propagation of the shell (see Section \ref{implications}). Coupling between SNe and AGN as well as geometry effects are other limitations to this simple approach, that are not adressed in this paper.

\section{Implications}
\label{implications}

The critical condition for gas removal is that the velocity $V_{\rm shell}$ of the shell of swept-up ISM is above the escape velocity when it reaches the virial radius. For both SNe and AGN, the scaling relations are such that this critical condition defines two mass regions on both sides of the critical halo mass: for halo masses below (respectively above) the critical halo mass, gas removal by the outflow is possible (impossible). 
We numerically integrate the equations of motion and compute the critical halo mass as the greatest for which the shell velocity exceeds the escape velocity at the virial radius. For AGN and SNe driven shells, we use the corresponding luminosity to compute the velocity of the shell, until it reaches the virial radius or stalls.  For SNe, we update the luminosity at each time step by updating the available gas mass: we remove the amount of gas that has gone to star formation from the total available gas mass.

\subsection{Cooling}
\label{cooling}
If the cooling timescale is shorter than the flow time, $R_{\rm shell}/V_{\rm shell}$, the cooling is efficient and the outflow is momentum driven, otherwise, it is energy-driven. We include cooling processes in our model when integrating the equation of motion of the blast wave, and choose between momentum- and energy-driven accordingly.
The dominant process for the cooling of the shocked wind is Compton cooling: the electrons of the shocked wind lose energy to photons via the inverse Compton effect  (\citealt{CO97}, \citealt{Hartwig17}). The Compton cooling time is given by

\begin{equation}
t_{\rm{ C}}= \frac{3 m_e c }{8 \pi \sigma_T U_{\rm{ rad}}} \frac{m_e c^2}{E}\ ,
\end{equation}

\noindent where $m_{e}$ is the mass of an electron, $E= 9 m_{\rm{ p}} v_{\rm{ w}}^2 /16$ is the energy of an electron in the shocked wind, and $ U_{\rm{ rad}} = L_{\rm{ AGN}}/(4 \pi R_{\rm sw}^2 c) $ is the radiation density in the wind.  At the beginning of the outflow, close to the IMBH, the radiation field is intense enough to cool the shocked wind: the outflow starts momentum driven, and $t_{ \rm C} \propto R^2$. Integrating equation \eqref{momentum}, the flow time in the momentum-driven regime follows $t_{\rm flow} \propto R^{2 + s/2}$, where $s<-1$ is the local slope of the NFW profile. Comparing the scaling of $t_{\rm C}(R)$ and $t_{\rm flow}(R)$, one sees that while gravity is negligible and as long as the AGN shines, a radius of transition to an energy-driven expansion exists.

To compute the effect of other cooling mechanisms, we use the \cite{SD93} cooling function approximated in polynomial form by \cite{TZ2001}. We find that subsequent radiative cooling becomes important only once the shell has slowed down to velocities much lower than the escape velocity and therefore has a negligible influence on the fate of the swept-up gas (see also \citealt{Hartwig17}). 

\begin{figure}
    \centering
    \includegraphics[width=\columnwidth]{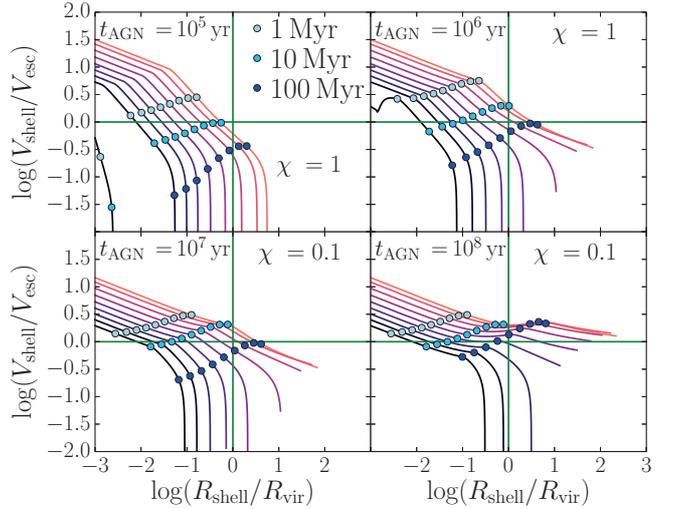}
    \caption{Velocity of the shell (normalized to the escape velocity at the virial radius) as a function of the radius of the shell (normalized to the virial radius), for different halo masses -- from $M_{\rm halo} = 10^5\rm{M}_{\odot}$ (top curve in each panel) up to $M_{\rm halo} = 10^{13}\rm{M}_{\odot}$ (bottom curve in each panel), with all the integer powers of ten in between. We assume an exponent $\alpha = 4$ in the $M_{\rm BH}-\sigma$ relation. The time markers give the propagation time after, $1$ Myr, $10$ Myr and $100$ Myr. We show the propagation for four pairs ($\chi$, $t_{\rm AGN}$). The critical halo mass is the greatest for which gas removal is possible, i.e. $V_{\rm shell}(R_{\rm vir}) > V_{\rm esc}( R_{\rm vir}) $.}
    \label{prop}
\end{figure}

Fig.~\ref{prop} displays the propagation for different ($\chi$, $t_{\rm AGN}$) pairs. The critical mass is the halo mass (1) below which the shell velocity is above the escape velocity at the virial radius, (2) above which the shell slows down before reaching the virial radius because of the inertia of the swept-up gas, gravity and/or cooling. The transition from the initial momentum- to energy-driven is visible in the upper-right panel: it corresponds to the abrupt early rise of the velocity of the bottom curve. For each panel, one can see a break in the velocity curve (e.g. at $t=t_{\rm AGN}=1\,\rm{Myr}$ in the upper-right panel), when the AGN stops shining. The smooth reacceleration of the shell, visible in the top curve of the lower-right panel, is due to the increasingly steeper slope of the NFW profile: beyond the scale radius, the gas is so tenuous that the shell can reaccelerate.
Note that we do not consider two-temperature effects \citep{FG2012}, which increases the Compton cooling time and thus leads to an earlier transition to adiabatic expansion, and therefore enhances gas ejection.

\subsection{Parameter study}
\label{precise}

\begin{figure*}
    \centering
    \includegraphics[width=\textwidth]{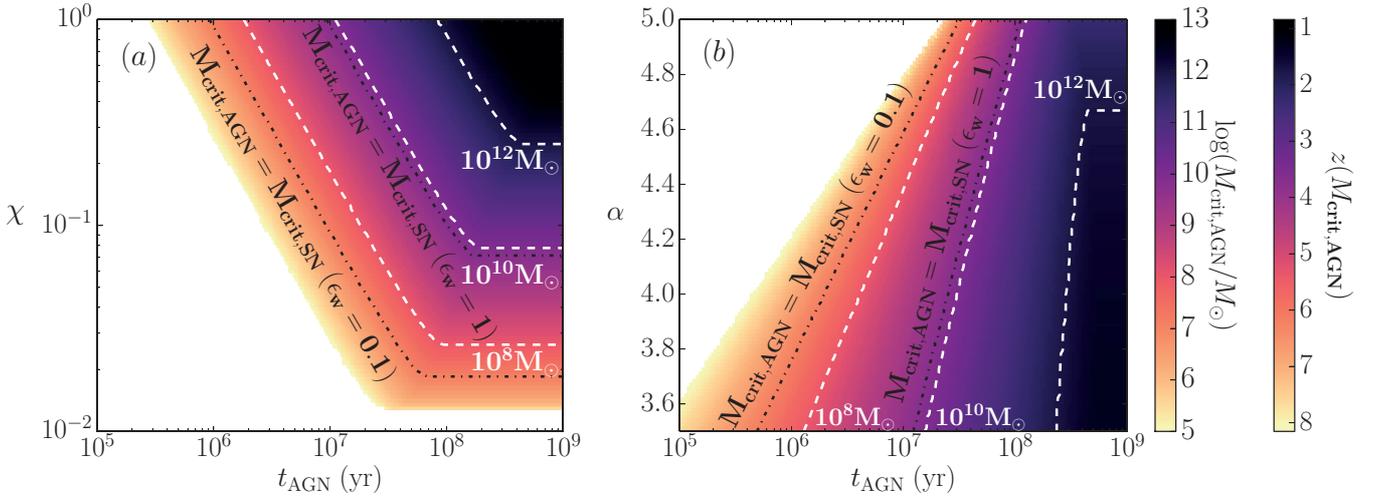}
    \caption{Critical halo mass below which gas removal by the AGN is possible, i.e. $V_{\rm shell}( R_{\rm vir}) > V_{\rm esc}( R_{\rm vir}) $, \textbf{(a)} as a function of the Eddington ratio $\chi$ and $t_{\rm AGN}$, with the slope of the $M_{\rm BH}- \sigma$ relation $\alpha=4$; \textbf{(b)} as a function of $\alpha$ and $t_{\rm AGN}$, with $\chi=0.3$. We show in a second colorbar the corresponding redshift according to equation \eqref{z}. White dashed contours indicate $M_{\rm{ crit}}=10^8 \rm{M}_{\odot}$, $10^{10} \rm{M}_{\odot}$ and $10^{12} \rm{M}_{\odot}$. In panel (a), the critical mass increases towards the upper-right corner of the panel. Therefore, for a given halo mass, gas removal is possible for the parameters $(t_{\rm AGN}, \chi)$ that are on the upper-right side of the corresponding white dashed contour. Similarly, in panel (b), the critical halo mass increases towards the lower-right side of the panel. Black dash-dotted contours show where the critical halo masses for AGN and SNe are equal. Note that the critical mass for SNe, at a given value of $\epsilon_{\rm w}$, is constant over these plots. Since a higher critical halo mass indicates stronger feedback, black contours also split the parameter space into two regions: in panel (a), AGN feedback dominates SN feedback in areas on the upper-right side of the black contours, and vice versa; in panel (b) AGN dominate in areas on the lower-right side of the black contours, and vice versa. AGN feedback is greater than SNe feedback in a significant part of the parameter space.}
    \label{crit_gravity}
\end{figure*}

Fig.~\ref{crit_gravity} shows the critical halo mass: for each pair of parameters $(t_{\rm AGN},\chi)$ (panel a) and $(t_{\rm AGN},\alpha)$ (panel b), we compute the critical halo mass below which gas removal by the AGN is possible, i.e. for which $V_{\rm shell}( R_{\rm vir}) > V_{\rm esc}( R_{\rm vir}) $. White dashed contours indicate $M_{\rm{ crit}}=10^8 \rm{M}_{\odot}$, $10^{10} \rm{M}_{\odot}$ and $10^{12} \rm{M}_{\odot}$. In panel (a), the critical mass increases towards the upper-right corner of the panel. Therefore, for a given halo mass, gas removal is possible for the parameters $(t_{\rm AGN}, \chi)$ that are on the upper-right side of the corresponding white dashed contour. Similarly, in panel (b), the critical halo mass increases towards the lower-right side of the panel. Black dash-dotted contours show where the critical halo masses for AGN and SNe are equal for given values of $\epsilon_{\rm w}$ ($0.1$ and $1$). Note that the critical mass for SNe, at a given value of $\epsilon_{\rm w}$, is constant over these plots. Since a higher critical halo mass indicates stronger feedback, these contours also split the parameter space into two regions: in panel (a), AGN feedback dominates SN feedback on the upper-right side of the black contours, SN feedback dominates AGN feedback on the lower-left side; in panel (b) AGN dominate on the lower-right side of the black contours, and vice-versa. AGN feedback is greater than SNe feedback in a significant part of the parameter space, since \cite{DT08} compute values of $\epsilon_{\rm w}$ smaller than 10\%. 

In panel (a) of Fig.~\ref{crit_gravity}, one sees that the contour plots for low halo masses follow straight lines with a slope of $-1$. This occurs because the outflow is mostly energy-driven for masses below the critical halo mass: what mostly counts is the input energy, i.e. $L_{\rm m} t_{\rm AGN}$, which is constant along lines of slope $-1$. However, the critical halo mass does not depend on $t_{\rm AGN}$ above a certain value of $t_{\rm AGN}$. The reason is that the energy needs to be applied on a timescale shorter than the free-fall time of the galaxy in order to overcome gravity. Regarding the comparison with SNe feedback: given a halo mass, and a value of $\epsilon_{\rm w}$, the SN luminosity is uniquely defined, which means that for parameters along the contours where $M_{\rm crit,SN} = M_{\rm crit,AGN}$, the luminosity for SNe is uniform (e.g. $L_{\rm{SN}} = 3\times 10^{40}$ erg, for $\epsilon_{\rm w} = 1$), whereas the AGN kinetic luminosity varies, and yet the critical mass is the same. The reason is that we turn off the AGN after $t_{\rm AGN}$, but do not turn off SN kinetic power. Therefore, input energy and kinetic luminosities are good first order approaches but one has to consider the timescales during which the energies are applied.

 In panel (b), one sees that the contour plots are closer for higher values of $\alpha$, which means that the dependence of the critical halo mass as a function of $t_{\rm AGN}$ at fixed $\alpha$ is steeper for higher values of $\alpha$. The reason is that the ratio $E_{\rm input} / E_{\rm grav}$, which quantifies to first order the effect of the AGN on the galaxy, is a shallower function of $M_{\rm halo}$ for higher values of $\alpha$, for which the value of the critical halo mass is more sensitive to the normalization of that ratio, and thus to $t_{\rm AGN}$.
Note that when computing the expansion of the SN wind, we do not consider the cooling of the hot interior of the bubble, which, if significant, should lower the critical halo mass for SNe and thus widen the subspace of AGN predominance.

\begin{figure}
    \centering
    \includegraphics[width=\columnwidth]{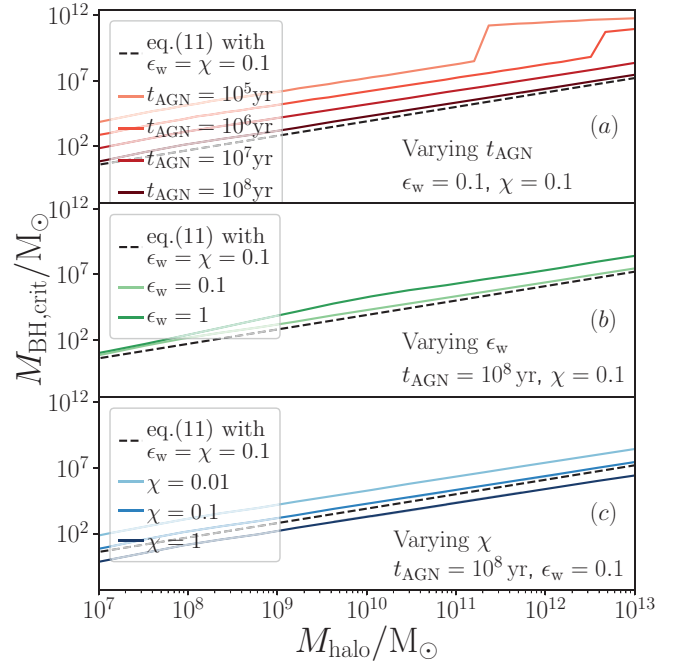}
	\caption{Critical black hole mass above which, for a given halo mass, AGN feedback is stronger than SN feedback for different values of \textbf{(a)} the duration of the AGN, $t_{\mathrm{AGN}}$; \textbf{(b)} the SN wind efficiency $\epsilon_{\mathrm{w}}$; \textbf{(c)} the Eddington ratio $\chi$. The black dashed line indicates the scaling found energetically in equation \eqref{mbhcrit}. The scalings found energetically and by solving the equation of motion of the shell are similar except for the short AGN time durations ($t_{\mathrm{AGN}}=10^5 \, \mathrm{yr}$ and $t_{\mathrm{AGN}}=10^6 \, \mathrm{yr}$) for which the sharp rise reveals the transition from an energy-driven to a momentum-driven only AGN outflow.}
	\label{eq11}    
    \end{figure}
    
In the manner of equation \eqref{mbhcrit}, we can release the constraint on the black hole mass given by the $M_{\mathrm{BH}}-\sigma$ relation and compute, using the equations of motions, the critical black hole mass above which AGN feedback is stronger than SN feedback. Fig.~\ref{eq11} shows the critical black hole mass above which the AGN-driven shell is pushed further in the ISM than the SN driven shell, as a function of the halo mass, for a given set of parameters $(t_{\mathrm{AGN}},\epsilon_{\mathrm{w}},\chi)$. One sees that the scalings found energetically in equation \eqref{mbhcrit} and by solving the equation of motion of the shell are similar. We retrieve the linear dependence of the critical black hole mass on $\epsilon_{\mathrm{w}}/\chi$, computed energetically in equation \eqref{mbhcrit}. However, in the upper panel, the influence of the duration of the AGN can be seen with the sharp rise in the critical BH-to-halo-mass relation, which does not appear in equation \eqref{mbhcrit}. This rise, for $t_{\mathrm{AGN}}=10^5 \, \mathrm{yr}$ and $t_{\mathrm{AGN}}=10^6 \, \mathrm{yr}$, indicates that the transition to an energy-driven propagation occurs later for higher black hole masses, and, hence, that transition never occurs for short $t_\mathrm{AGN}$ and high black hole masses.

\section{Discussion and conclusion}
\label{discussion}

In this paper, we investigated the possibility of AGN feedback in dwarf galaxies. Assuming scaling relations between the relevant physical quantities, we obtained a critical halo mass below which gas removal by the AGN is possible. In a broad part of the parameter space, AGN feedback is more efficient than SNe feedback. This suggests that AGN could succeed where SNe might fail, such as in the most massive dwarf galaxies. We argue that AGN could potentially have played a significant role in gas ejection in early dwarf evolution.

In our definition, AGN feedback is \textit{efficient} when the AGN-driven shell escapes the virial radius of the galaxy. This does not mean that AGN feedback is not efficient at regulating star formation in massive halos, just that there are no escaping winds. In particular, we do not include the effect of AGN feedback on the accretion of cooling flows. Moreover, in high redshift small mass disc galaxies, the effect of AGN should be smaller than in our 1D model because the outflow can escape in the perpendicular direction as analyzed by \cite{Hartwig17}. Besides, our work does not address the interplay between SN and AGN feedback: efficient SN feedback can prevent the accumulation of dense cold gas and starve the black hole \citep{DV2015}. A full treatment of this phenomenon would constrain the explorable parameter space for the black hole and minimize the predominance of AGN. A complete treatment would also include a multiphase ISM, which, when taken into consideration, reduces the feedback efficiency for both AGN \citep{Costa14} and SNe \citep{BH15}. AGN can also trigger star formation as proposed theoretically \citep{SN09,Gaibler12} and seen observationally \citep{Maiolino17}. More realistic simulations, including much of this additional physics, are needed before we can fully understand the role of AGN feedback in the multiphase ISM of initially gas-rich dwarf galaxies.

\section*{Acknowledgements}
We thank the anonymous referee for useful remarks. We are grateful to Marta Volonteri for fruitful discussions. The work of JS has been supported in part by ERC
Project No. 267117 (DARK) hosted by the Pierre and
Marie Curie University-Paris VI, Sorbonne Universities
and CEA-Saclay.
TH acknowledges funding under the European Community's Seventh Framework Programme (FP7/2007-2013) via the European Research Council Grants `BLACK' under the project number 614199.


\bibliographystyle{mnras}
\bibliography{bibliography} 

\begin{thebibliography}{}
\makeatletter
\relax
\def\mn@urlcharsother{\let\do\@makeother \do\$\do\&\do\#\do\^\do\_\do\%\do\~}
\def\mn@doi{\begingroup\mn@urlcharsother \@ifnextchar [ {\mn@doi@}
  {\mn@doi@[]}}
\def\mn@doi@[#1]#2{\def\@tempa{#1}\ifx\@tempa\@empty \href
  {http://dx.doi.org/#2} {doi:#2}\else \href {http://dx.doi.org/#2} {#1}\fi
  \endgroup}
\def\mn@eprint#1#2{\mn@eprint@#1:#2::\@nil}
\def\mn@eprint@arXiv#1{\href {http://arxiv.org/abs/#1} {{\tt arXiv:#1}}}
\def\mn@eprint@dblp#1{\href {http://dblp.uni-trier.de/rec/bibtex/#1.xml}
  {dblp:#1}}
\def\mn@eprint@#1:#2:#3:#4\@nil{\def\@tempa {#1}\def\@tempb {#2}\def\@tempc
  {#3}\ifx \@tempc \@empty \let \@tempc \@tempb \let \@tempb \@tempa \fi \ifx
  \@tempb \@empty \def\@tempb {arXiv}\fi \@ifundefined
  {mn@eprint@\@tempb}{\@tempb:\@tempc}{\expandafter \expandafter \csname
  mn@eprint@\@tempb\endcsname \expandafter{\@tempc}}}

\bibitem[\protect\citeauthoryear{{Baldassare}, {Reines}, {Gallo}  \&
  {Greene}}{{Baldassare} et~al.}{2017}]{Baldassare2016}
{Baldassare} V.~F.,  {Reines} A.~E.,  {Gallo} E.,   {Greene} J.~E.,  2017,
  \mn@doi [\apj] {10.3847/1538-4357/836/1/20}, \href
  {http://adsabs.harvard.edu/abs/2017ApJ...836...20B} {836, 20}

\bibitem[\protect\citeauthoryear{{Bland-Hawthorn}, {Sutherland}  \&
  {Webster}}{{Bland-Hawthorn} et~al.}{2015}]{BH15}
{Bland-Hawthorn} J.,  {Sutherland} R.,   {Webster} D.,  2015, \mn@doi [\apj]
  {10.1088/0004-637X/807/2/154}, \href
  {http://adsabs.harvard.edu/abs/2015ApJ...807..154B} {807, 154}

\bibitem[\protect\citeauthoryear{{Bower}, {Benson}, {Malbon}, {Helly}, {Frenk},
  {Baugh}, {Cole}  \& {Lacey}}{{Bower} et~al.}{2006}]{Bower2006}
{Bower} R.~G.,  {Benson} A.~J.,  {Malbon} R.,  {Helly} J.~C.,  {Frenk} C.~S.,
  {Baugh} C.~M.,  {Cole} S.,   {Lacey} C.~G.,  2006, \mn@doi [\mnras]
  {10.1111/j.1365-2966.2006.10519.x}, \href
  {http://adsabs.harvard.edu/abs/2006MNRAS.370..645B} {370, 645}

\bibitem[\protect\citeauthoryear{{Boylan-Kolchin}, {Bullock}  \&
  {Kaplinghat}}{{Boylan-Kolchin} et~al.}{2011}]{BK11}
{Boylan-Kolchin} M.,  {Bullock} J.~S.,   {Kaplinghat} M.,  2011, \mn@doi
  [\mnras] {10.1111/j.1745-3933.2011.01074.x}, \href
  {http://adsabs.harvard.edu/abs/2011MNRAS.415L..40B} {415, L40}

\bibitem[\protect\citeauthoryear{{Bryan} \& {Norman}}{{Bryan} \&
  {Norman}}{1998}]{BN98}
{Bryan} G.~L.,  {Norman} M.~L.,  1998, \mn@doi [\apj] {10.1086/305262}, \href
  {http://adsabs.harvard.edu/abs/1998ApJ...495...80B} {495, 80}

\bibitem[\protect\citeauthoryear{{Ciotti} \& {Ostriker}}{{Ciotti} \&
  {Ostriker}}{1997}]{CO97}
{Ciotti} L.,  {Ostriker} J.~P.,  1997, \mn@doi [\apjl] {10.1086/310902}, \href
  {http://adsabs.harvard.edu/abs/1997ApJ...487L.105C} {487, L105}

\bibitem[\protect\citeauthoryear{{Cole} \& {Lacey}}{{Cole} \&
  {Lacey}}{1996}]{CL96}
{Cole} S.,  {Lacey} C.,  1996, \mn@doi [\mnras] {10.1093/mnras/281.2.716},
  \href {http://adsabs.harvard.edu/abs/1996MNRAS.281..716C} {281, 716}

\bibitem[\protect\citeauthoryear{{Costa}, {Sijacki}  \& {Haehnelt}}{{Costa}
  et~al.}{2014}]{Costa14}
{Costa} T.,  {Sijacki} D.,   {Haehnelt} M.~G.,  2014, \mn@doi [\mnras]
  {10.1093/mnras/stu1632}, \href
  {http://adsabs.harvard.edu/abs/2014MNRAS.444.2355C} {444, 2355}

\bibitem[\protect\citeauthoryear{{Croton} et~al.,}{{Croton}
  et~al.}{2006}]{Croton2006}
{Croton} D.~J.,  et~al., 2006, \mn@doi [\mnras]
  {10.1111/j.1365-2966.2005.09675.x}, \href
  {http://adsabs.harvard.edu/abs/2006MNRAS.365...11C} {365, 11}

\bibitem[\protect\citeauthoryear{{Dekel} \& {Silk}}{{Dekel} \&
  {Silk}}{1986}]{DS86}
{Dekel} A.,  {Silk} J.,  1986, \mn@doi [\apj] {10.1086/164050}, \href
  {http://adsabs.harvard.edu/abs/1986ApJ...303...39D} {303, 39}

\bibitem[\protect\citeauthoryear{{Dubois} \& {Teyssier}}{{Dubois} \&
  {Teyssier}}{2008}]{DT08}
{Dubois} Y.,  {Teyssier} R.,  2008, \mn@doi [\aap]
  {10.1051/0004-6361:20078326}, \href
  {http://adsabs.harvard.edu/abs/2008A%26A...477...79D} {477, 79}

\bibitem[\protect\citeauthoryear{{Dubois}, {Volonteri}, {Silk}, {Devriendt},
  {Slyz}  \& {Teyssier}}{{Dubois} et~al.}{2015}]{DV2015}
{Dubois} Y.,  {Volonteri} M.,  {Silk} J.,  {Devriendt} J.,  {Slyz} A.,
  {Teyssier} R.,  2015, \mn@doi [\mnras] {10.1093/mnras/stv1416}, \href
  {http://adsabs.harvard.edu/abs/2015MNRAS.452.1502D} {452, 1502}

\bibitem[\protect\citeauthoryear{{Dutton} \& {Macci{\`o}}}{{Dutton} \&
  {Macci{\`o}}}{2014}]{DM14}
{Dutton} A.~A.,  {Macci{\`o}} A.~V.,  2014, \mn@doi [\mnras]
  {10.1093/mnras/stu742}, \href
  {http://adsabs.harvard.edu/abs/2014MNRAS.441.3359D} {441, 3359}

\bibitem[\protect\citeauthoryear{{Faucher-Gigu{\`e}re} \&
  {Quataert}}{{Faucher-Gigu{\`e}re} \& {Quataert}}{2012}]{FG2012}
{Faucher-Gigu{\`e}re} C.-A.,  {Quataert} E.,  2012, \mn@doi [\mnras]
  {10.1111/j.1365-2966.2012.21512.x}, \href
  {http://adsabs.harvard.edu/abs/2012MNRAS.425..605F} {425, 605}

\bibitem[\protect\citeauthoryear{{Ferrarese} \& {Merritt}}{{Ferrarese} \&
  {Merritt}}{2000}]{FM2000}
{Ferrarese} L.,  {Merritt} D.,  2000, \mn@doi [\apjl] {10.1086/312838}, \href
  {http://adsabs.harvard.edu/abs/2000ApJ...539L...9F} {539, L9}

\bibitem[\protect\citeauthoryear{{Gaibler}, {Khochfar}, {Krause}  \&
  {Silk}}{{Gaibler} et~al.}{2012}]{Gaibler12}
{Gaibler} V.,  {Khochfar} S.,  {Krause} M.,   {Silk} J.,  2012, \mn@doi
  [\mnras] {10.1111/j.1365-2966.2012.21479.x}, \href
  {http://adsabs.harvard.edu/abs/2012MNRAS.425..438G} {425, 438}

\bibitem[\protect\citeauthoryear{{Garrison-Kimmel}, {Rocha}, {Boylan-Kolchin},
  {Bullock}  \& {Lally}}{{Garrison-Kimmel} et~al.}{2013}]{GK13}
{Garrison-Kimmel} S.,  {Rocha} M.,  {Boylan-Kolchin} M.,  {Bullock} J.~S.,
  {Lally} J.,  2013, \mn@doi [\mnras] {10.1093/mnras/stt984}, \href
  {http://adsabs.harvard.edu/abs/2013MNRAS.433.3539G} {433, 3539}

\bibitem[\protect\citeauthoryear{{Garrison-Kimmel}, {Boylan-Kolchin}, {Bullock}
   \& {Kirby}}{{Garrison-Kimmel} et~al.}{2014}]{GK14}
{Garrison-Kimmel} S.,  {Boylan-Kolchin} M.,  {Bullock} J.~S.,   {Kirby} E.~N.,
  2014, \mn@doi [\mnras] {10.1093/mnras/stu1477}, \href
  {http://adsabs.harvard.edu/abs/2014MNRAS.444..222G} {444, 222}

\bibitem[\protect\citeauthoryear{{Gebhardt} et~al.,}{{Gebhardt}
  et~al.}{2000}]{Gebhardt2000}
{Gebhardt} K.,  et~al., 2000, \mn@doi [\apjl] {10.1086/312840}, \href
  {http://adsabs.harvard.edu/abs/2000ApJ...539L..13G} {539, L13}

\bibitem[\protect\citeauthoryear{{Genzel} et~al.,}{{Genzel}
  et~al.}{2015}]{Genzel2015}
{Genzel} R.,  et~al., 2015, \mn@doi [\apj] {10.1088/0004-637X/800/1/20}, \href
  {http://cdsads.u-strasbg.fr/abs/2015ApJ...800...20G} {800, 20}

\bibitem[\protect\citeauthoryear{{Gofford}, {Reeves}, {Tombesi}, {Braito},
  {Turner}, {Miller}  \& {Cappi}}{{Gofford} et~al.}{2013}]{Gofford2013}
{Gofford} J.,  {Reeves} J.~N.,  {Tombesi} F.,  {Braito} V.,  {Turner} T.~J.,
  {Miller} L.,   {Cappi} M.,  2013, \mn@doi [\mnras] {10.1093/mnras/sts481},
  \href {http://adsabs.harvard.edu/abs/2013MNRAS.430...60G} {430, 60}

\bibitem[\protect\citeauthoryear{{G{\"u}ltekin} et~al.,}{{G{\"u}ltekin}
  et~al.}{2009}]{Gultekin2009}
{G{\"u}ltekin} K.,  et~al., 2009, \mn@doi [\apj] {10.1088/0004-637X/698/1/198},
  \href {http://adsabs.harvard.edu/abs/2009ApJ...698..198G} {698, 198}

\bibitem[\protect\citeauthoryear{{Hartwig}, {Volonteri}  \&
  {Dashyan}}{{Hartwig} et~al.}{2017}]{Hartwig17}
{Hartwig} T.,  {Volonteri} M.,   {Dashyan} G.,  2017, preprint, \href
  {http://adsabs.harvard.edu/abs/2017arXiv170703826H} {} (\mn@eprint {arXiv}
  {1707.03826})

\bibitem[\protect\citeauthoryear{{Kim} \& {Ostriker}}{{Kim} \&
  {Ostriker}}{2015}]{KO15}
{Kim} C.-G.,  {Ostriker} E.~C.,  2015, \mn@doi [\apj]
  {10.1088/0004-637X/802/2/99}, \href
  {http://adsabs.harvard.edu/abs/2015ApJ...802...99K} {802, 99}

\bibitem[\protect\citeauthoryear{{Kim}, {Ostriker}  \& {Raileanu}}{{Kim}
  et~al.}{2017}]{KO17}
{Kim} C.-G.,  {Ostriker} E.~C.,   {Raileanu} R.,  2017, \mn@doi [\apj]
  {10.3847/1538-4357/834/1/25}, \href
  {http://adsabs.harvard.edu/abs/2017ApJ...834...25K} {834, 25}

\bibitem[\protect\citeauthoryear{{King} \& {Pounds}}{{King} \&
  {Pounds}}{2015}]{KP15}
{King} A.,  {Pounds} K.,  2015, \mn@doi [\araa]
  {10.1146/annurev-astro-082214-122316}, \href
  {http://adsabs.harvard.edu/abs/2015ARA%26A..53..115K} {53, 115}

\bibitem[\protect\citeauthoryear{{Koo} \& {McKee}}{{Koo} \&
  {McKee}}{1990}]{KM92}
{Koo} B.-C.,  {McKee} C.~F.,  1990, \mn@doi [\apj] {10.1086/168712}, \href
  {http://adsabs.harvard.edu/abs/1990ApJ...354..513K} {354, 513}

\bibitem[\protect\citeauthoryear{{Kormendy} \& {Ho}}{{Kormendy} \&
  {Ho}}{2013}]{KH13}
{Kormendy} J.,  {Ho} L.~C.,  2013, \mn@doi [\araa]
  {10.1146/annurev-astro-082708-101811}, \href
  {http://adsabs.harvard.edu/abs/2013ARA%26A..51..511K} {51, 511}

\bibitem[\protect\citeauthoryear{{Kroupa}}{{Kroupa}}{2001}]{Kroupa}
{Kroupa} P.,  2001, \mn@doi [\mnras] {10.1046/j.1365-8711.2001.04022.x}, \href
  {http://adsabs.harvard.edu/abs/2001MNRAS.322..231K} {322, 231}

\bibitem[\protect\citeauthoryear{{{\L}okas} \& {Mamon}}{{{\L}okas} \&
  {Mamon}}{2001}]{LM01}
{{\L}okas} E.~L.,  {Mamon} G.~A.,  2001, \mn@doi [\mnras]
  {10.1046/j.1365-8711.2001.04007.x}, \href
  {http://adsabs.harvard.edu/abs/2001MNRAS.321..155L} {321, 155}

\bibitem[\protect\citeauthoryear{{Maiolino} et~al.,}{{Maiolino}
  et~al.}{2017}]{Maiolino17}
{Maiolino} R.,  et~al., 2017, \mn@doi [\nat] {10.1038/nature21677}, \href
  {http://adsabs.harvard.edu/abs/2017Natur.544..202M} {544, 202}

\bibitem[\protect\citeauthoryear{{Miller}, {Gallo}, {Greene}, {Kelly}, {Treu},
  {Woo}  \& {Baldassare}}{{Miller} et~al.}{2015}]{Miller15}
{Miller} B.~P.,  {Gallo} E.,  {Greene} J.~E.,  {Kelly} B.~C.,  {Treu} T.,
  {Woo} J.-H.,   {Baldassare} V.,  2015, \mn@doi [\apj]
  {10.1088/0004-637X/799/1/98}, \href
  {http://adsabs.harvard.edu/abs/2015ApJ...799...98M} {799, 98}

\bibitem[\protect\citeauthoryear{{Moore}, {Ghigna}, {Governato}, {Lake},
  {Quinn}, {Stadel}  \& {Tozzi}}{{Moore} et~al.}{1999}]{Moore99}
{Moore} B.,  {Ghigna} S.,  {Governato} F.,  {Lake} G.,  {Quinn} T.,  {Stadel}
  J.,   {Tozzi} P.,  1999, \mn@doi [\apjl] {10.1086/312287}, \href
  {http://adsabs.harvard.edu/abs/1999ApJ...524L..19M} {524, L19}

\bibitem[\protect\citeauthoryear{{Navarro}, {Frenk}  \& {White}}{{Navarro}
  et~al.}{1996}]{NFW}
{Navarro} J.~F.,  {Frenk} C.~S.,   {White} S.~D.~M.,  1996, \mn@doi [\apj]
  {10.1086/177173}, \href {http://adsabs.harvard.edu/abs/1996ApJ...462..563N}
  {462, 563}

\bibitem[\protect\citeauthoryear{{Oh}, {Brook}, {Governato}, {Brinks}, {Mayer},
  {de Blok}, {Brooks}  \& {Walter}}{{Oh} et~al.}{2011}]{Oh11}
{Oh} S.-H.,  {Brook} C.,  {Governato} F.,  {Brinks} E.,  {Mayer} L.,  {de Blok}
  W.~J.~G.,  {Brooks} A.,   {Walter} F.,  2011, \mn@doi [\aj]
  {10.1088/0004-6256/142/1/24}, \href
  {http://adsabs.harvard.edu/abs/2011AJ....142...24O} {142, 24}

\bibitem[\protect\citeauthoryear{{Pardo} et~al.,}{{Pardo}
  et~al.}{2016}]{Pardo2016}
{Pardo} K.,  et~al., 2016, \mn@doi [\apj] {10.3847/0004-637X/831/2/203}, \href
  {http://adsabs.harvard.edu/abs/2016ApJ...831..203P} {831, 203}

\bibitem[\protect\citeauthoryear{{Silk}}{{Silk}}{2017}]{Silk_feedback}
{Silk} J.,  2017, \mn@doi [\apjl] {10.3847/2041-8213/aa67da}, \href
  {http://adsabs.harvard.edu/abs/2017ApJ...839L..13S} {839, L13}

\bibitem[\protect\citeauthoryear{{Silk} \& {Norman}}{{Silk} \&
  {Norman}}{2009}]{SN09}
{Silk} J.,  {Norman} C.,  2009, \mn@doi [\apj] {10.1088/0004-637X/700/1/262},
  \href {http://adsabs.harvard.edu/abs/2009ApJ...700..262S} {700, 262}

\bibitem[\protect\citeauthoryear{{Sutherland} \& {Dopita}}{{Sutherland} \&
  {Dopita}}{1993}]{SD93}
{Sutherland} R.~S.,  {Dopita} M.~A.,  1993, \mn@doi [\apjs] {10.1086/191823},
  \href {http://adsabs.harvard.edu/abs/1993ApJS...88..253S} {88, 253}

\bibitem[\protect\citeauthoryear{{Tombesi}, {Cappi}, {Reeves}, {Palumbo},
  {Yaqoob}, {Braito}  \& {Dadina}}{{Tombesi} et~al.}{2010}]{Tombesi2010}
{Tombesi} F.,  {Cappi} M.,  {Reeves} J.~N.,  {Palumbo} G.~G.~C.,  {Yaqoob} T.,
  {Braito} V.,   {Dadina} M.,  2010, \mn@doi [\aap]
  {10.1051/0004-6361/200913440}, \href
  {http://adsabs.harvard.edu/abs/2010A%26A...521A..57T} {521, A57}

\bibitem[\protect\citeauthoryear{{Tozzi} \& {Norman}}{{Tozzi} \&
  {Norman}}{2001}]{TZ2001}
{Tozzi} P.,  {Norman} C.,  2001, \mn@doi [\apj] {10.1086/318237}, \href
  {http://adsabs.harvard.edu/abs/2001ApJ...546...63T} {546, 63}

\bibitem[\protect\citeauthoryear{{Tremaine} et~al.,}{{Tremaine}
  et~al.}{2002}]{Tremaine2000}
{Tremaine} S.,  et~al., 2002, \mn@doi [\apj] {10.1086/341002}, \href
  {http://adsabs.harvard.edu/abs/2002ApJ...574..740T} {574, 740}

\bibitem[\protect\citeauthoryear{{Weaver}, {McCray}, {Castor}, {Shapiro}  \&
  {Moore}}{{Weaver} et~al.}{1977}]{weaver77}
{Weaver} R.,  {McCray} R.,  {Castor} J.,  {Shapiro} P.,   {Moore} R.,  1977,
  \mn@doi [\apj] {10.1086/155692}, \href
  {http://adsabs.harvard.edu/abs/1977ApJ...218..377W} {218, 377}

\bibitem[\protect\citeauthoryear{{van den Bosch}}{{van den
  Bosch}}{2002}]{VDB02}
{van den Bosch} F.~C.,  2002, \mn@doi [\mnras]
  {10.1046/j.1365-8711.2002.05171.x}, \href
  {http://adsabs.harvard.edu/abs/2002MNRAS.331...98V} {331, 98}

\makeatother
\end{thebibliography}

\bsp	
\label{lastpage}
\end{document}